\documentclass[aps,prl,preprint,superscriptaddress]{revtex4-1}
\usepackage{gensymb}
\usepackage{graphicx}
\usepackage{dcolumn}
\usepackage{bm}
\usepackage{siunitx}
\usepackage{hyperref}
\usepackage[capitalise]{cleveref}
\usepackage[colorinlistoftodos]{todonotes}
\usepackage{float}
\usepackage{booktabs}
\newcommand{\ve}{\boldsymbol}
\usepackage{cleveref}
\usepackage{import}
\usepackage{afterpage}
\usepackage{gensymb}

\begin{document}

\title{Atomic scale control of spin current transmission at interfaces}

\author{Mohamed Amine Wahada}
\email{awahada@mpi-halle.mpg.de}
\affiliation{ Max Planck Institute for Microstructure Physics, Weinberg 2 06120 Halle, Germany}

\author{Ersoy Şaşıoğlu}
\affiliation{ Institute of Physics, Martin Luther University Halle-Wittenberg, Von-Seckendorff-Platz 1, 06120 Halle, Germany}

\author{Wolfgang Hoppe}
\affiliation{ Institute of Physics, Martin Luther University Halle-Wittenberg, von Danckelmann Platz 3, 06120 Halle, Germany}

\author{Xilin Zhou}
\affiliation{ Max Planck Institute for Microstructure Physics, Weinberg 2 06120 Halle, Germany}

\author{Hakan Deniz}
\affiliation{ Max Planck Institute for Microstructure Physics, Weinberg 2 06120 Halle, Germany}

\author{Reza Rouzegar}
\affiliation{ Institute of Physics, Freie Universität Berlin, Arnimalee 14, 01120 Berlin, Germany}

\author{Tobias Kampfrath}
\affiliation{ Institute of Physics, Freie Universität Berlin, Arnimalee 14, 01120 Berlin, Germany}

\author{Ingrid Mertig}
\affiliation{ Institute of Physics, Martin Luther University Halle-Wittenberg, Von-Seckendorff-Platz 1, 06120 Halle, Germany}

\author{Stuart S. P. Parkin}
\affiliation{ Max Planck Institute for Microstructure Physics, Weinberg 2 06120 Halle, Germany}

\author{Georg Woltersdorf}
 \email{georg.woltersdorf@physik.uni-halle.de}
\affiliation{ Institute of Physics, Martin Luther University Halle-Wittenberg, von Danckelmann Platz 3, 06120 Halle, Germany}

\begin{abstract}
Spin transmission at ferromagnet/heavy metal interfaces is of vital importance for many spintronic devices. Usually the spin current transmission is limited by the spin mixing conductance and loss mechanisms such as spin memory loss. In order to understand these effects, we study the interface transmission when an insulating interlayer is inserted between the ferromagnet and the heavy metal. For this we measure the inverse spin Hall voltage generated from optically injected spin current pulses as well as the magnitude of the spin pumping using ferromagnetic resonance. From our results we conclude that significant spin memory loss only occurs for 5\textit{d} metals with less than half filled \textit{d}-shell.

\end{abstract}

\pacs{Valid PACS appear here}%
\keywords{Spin Hall effect, ultrafast demagnetization}%
\maketitle

\section{Introduction}

Spin current  is the net flow of angular momentum and  in condensed matter usually carried by the spin degree of freedom of electrons\cite{Maekawa_OxfordUniversityPress_2012}. The discovery that the magnetization vector can be controlled in nanostuctures using spin currents \cite{Miron_Nature_2011,MihaiMiron_NatMat_2010,Liu_Science_2012,Liu_PRL_2012} provided an enormous boost to the field of spintronics. This effect is essential for spin based non volatile memory applications, displacement of spin textures such as chiral domain walls and skyrmions for racetrack applications\cite{Parkin_science_2008,Ryu_NatNanotech_2013, Fert_Natnanotech_2013,Iwasaki_Natnanotech_2013,Sampaio_Natnanotech_2013}, or spin torque oscillators \cite{Demidov_Natmat_2012}. 

A spin polarized current may be produced electrically  by passing a current in a ferromagnet (FM). Alternatively an unpolarized charge current can be used to generate  spin polarization at the surface of a non magnetic heavy metal (HM) with spin orbit coupling (SOC) via the spin Hall effect (SHE) \cite{Hoffmann_IEEE_2013,saitoh_apl_2006,Sinova_rev.mod.phys_2015}. The efficiency of this conversion is a material dependent property and usually quantified by the spin Hall angle. When interfaced with a FM, the spin current is transmitted through the interface and can exert a torque on the magnetization, a phenomenon known as spin orbit torque (SOT). Reciprocally, spins can be pumped from the FM into the HM where the SC is converted to a charge current via the inverse SHE (ISHE). This may be demonstrated via microwave induced precession of the magnetization \cite{saitoh_apl_2006,ando_jap_2011} or via ultrafast optical excitation \cite{Kampfrath_nanotech_2013,Seifert_natphotonics_2016}. The latter example is the basis of ultra broadband and efficient THz emitters \cite{Seifert_natphotonics_2016, Gueckstock_Adv_Mat_2021, Hoppe_ACS_anm_2021}.

The key element for these processes is the FM/HM interface\cite{Zhang_np_2015}. Mainly two effects can impair the spin current transmission. (i) the magnitude of the spin mixing conductance \cite{Xia-PRB2002} and (ii) spin memory loss (SML) \cite{Gupta_prl_2020,Rojas_prl_2014,Belashchenko_prl_2016,Dolui_prb_2017}.  SML has first been considered in magnetoresistance experiments\cite{Bass_iop_2007,Galinon_apl_2005}. 
The possible underlying mechanisms for SML are still debated, nevertheless it is believed to be related to interface spin orbit coupling (i-SOC) \cite{Zhu_prl_2019_SP,Zhu_prl_2019_ST,Flores_prb_2020,Dolui_prb_2017,Zeng_apl_2019}, non collinear magnetization, disorder \cite{Gupta_prl_2020,Wesselink_prb_2019}, and lattice mismatch \cite{Gupta_prl_2020}. SML due to i-SOC is often evoked in order to explain inconsistencies of the value of the spin Hall angle in spin transport experiments\cite{Tao_Scienceadv_2018,Yu_prm_2018,Chen_IEEE_2015,Chen_prl_2015}.
Separating, understanding, and finally controlling these interface properties is an essential step to enhance the spin current injection efficiency. Up to now only a few studies have been performed in this regard \cite{Zhu_prb_2019, Gueckstock_Adv_Mat_2021} (interface alloying) and \cite{Zhang_np_2015,Deorani_apl_2013, Panda_ScieAdv2019}(interlayer insertion). For the giant magnetoresistance samples the insertion of interface dusting layers lead to a dramatic enhancement of the magnetoresistance \cite{Parkin_apl_1992,Parkin_prl_1993}. In the same spirit, using non magnetic dusting layers between FM/HM allows to increase the efficiency of current induced chiral domain wall motion \cite{Guan_AdvancedMat_2021}. 

In this work we use MgO as an interlayer with varying thicknesses placed at the HM/FM interface. In order to identify the SML processes two experiments are used to independently determine (i) the total amount of spin current injected by the ferromagnet and (ii) the amount of spin current delivered into to the HM. 
By performing optical ultrafast spin injection and ISHE charge current detection, we demonstrate that ultrathin MgO layers reduce the spin current transmission differently depending on the HM element. While we find a reduction for HM=Pt, it is rather enhanced for HM=Ta. By the additional measurement of the total spin momentum pumped by the FM across the interface (spin pumping) we estimate the magnitude of SML. Using the MgO interlayer thickness dependence and corresponding calculations of the electronic structure we link the SML to the spin moment reduction at the FM/Ta interface.

\section{Results}
\begin{figure*}
\includegraphics[width=1\textwidth]{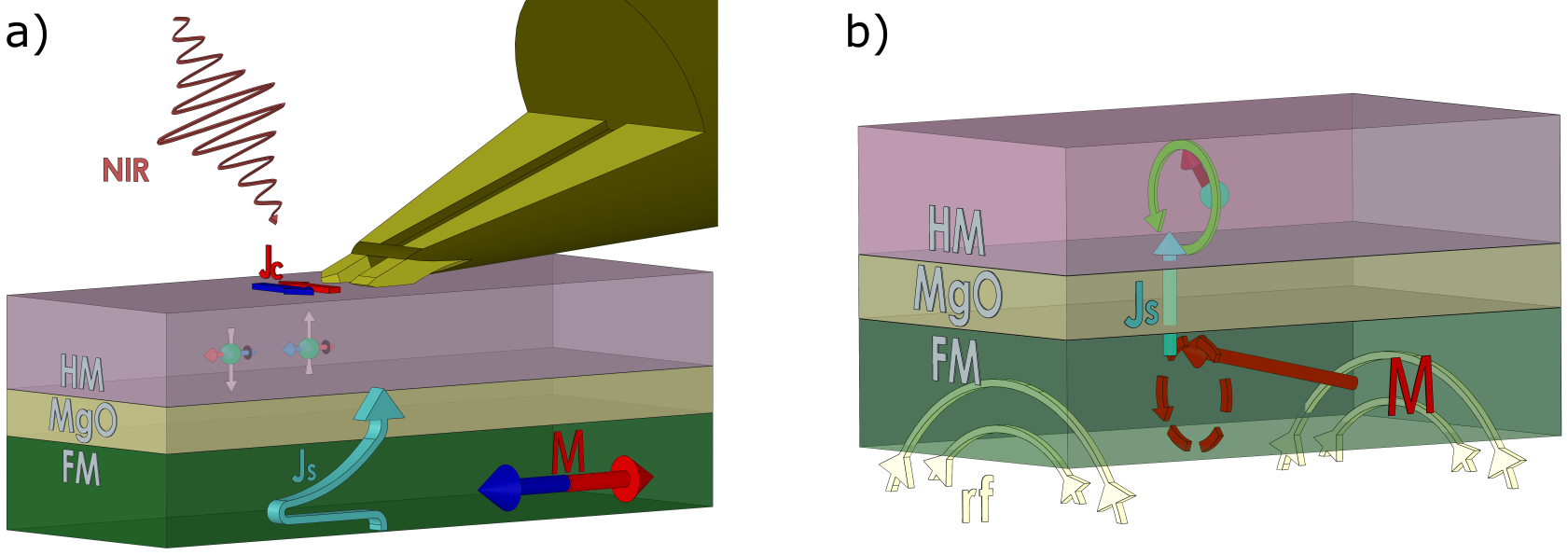}
\caption{\label{fig:1} {\bf Experimental configuration.} (a) Ultrafast measurement of ISHE using optical excitation and electronic detection. A pulsed femtosecond laser excites the HM/FM bilayer and creates an ultrafast spin current pulse $\ve{J}_s$ prapagating across the MgO interlayer into the  HM where it is converted into a charge current $\ve{J}_c$ picked up by an rf-probe tip.  Once $\ve{J}_s$ is converted in the HM, $\ve{J}_c$. (b) In a complementary spin pumping experiment, a continuous rf field excites the magnetization causing its precession which generates a spin current that is pumped from the ferromagnet to the heavy  metal. The spin transport into the HM enhances the damping in FM.} 
\end{figure*}
\begin{figure*}
\includegraphics[width=1\textwidth]{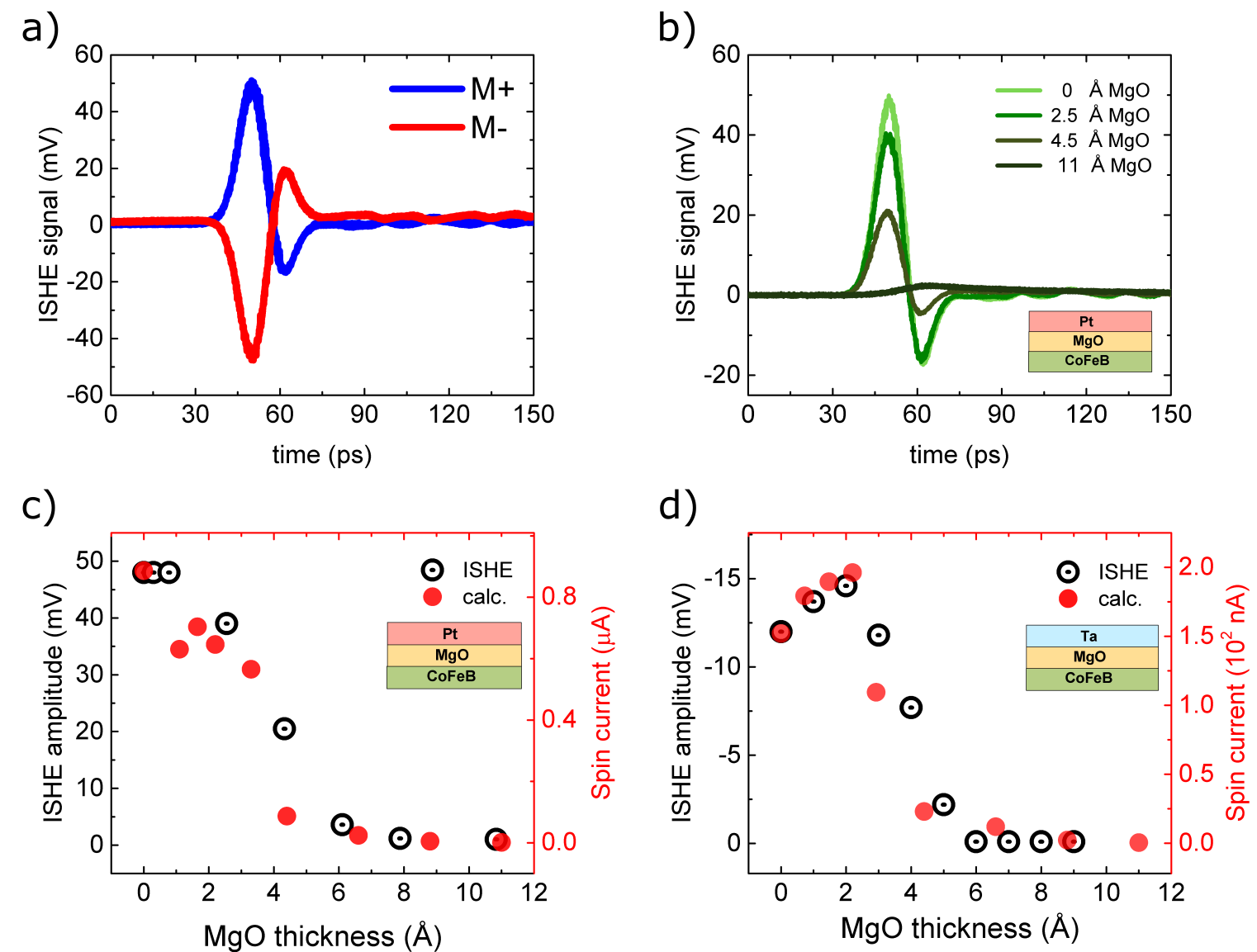}
\caption{\label{fig:2} {\bf Ultrafast ISHE measurements.} (a) Time resolved ISHE signal for both magnetization directions. (b) Time resolved ISHE signal for the HM=Pt sample with different MgO thicknesses ranging from 0 to \SI{17}{\AA}.(c)-(d) MgO thickness dependence of the ISHE signal amplitude and the calculated spin current transmission for the HM=Pt and HM =Ta, respectively. Note, that for HM=Ta, the ISHE signal has negative values due to the negative sign of the spin Hall angle of Ta. } 
\end{figure*}

{\bf Ultrafast measurements of the inverse SHE.}
Ultrafast spin current pulses are generated by exciting the layer stack with optical femtosecond laser pulses. The inverse SHE converts these spin current pulses into charge currents which are typically detected in the form of free space THz radiation using electro-optic sampling \cite{Seifert_natphotonics_2016}. Here we use coplanar probe tip to detect the sub-THz current pulse directly on the sample as shown in \cref{fig:1}(a). Typical results are shown in \cref{fig:2}(a). 

{\bf Impact of MgO interlayers.}
As a next step, we examine the influence of ultrathin MgO interlayers on the inverse SHE signals for HM=Pt. Different waveforms are shown in \cref{fig:2}(b) for various MgO thicknesses. The amplitudes of the signals are normalized to the peak signal without MgO interlayer and displayed in \cref{fig:2}(c). As a function of the MgO thickness one observes a plateau-like behavior for up to \SI{2}{\AA} of the MgO layer corresponding to the thickness of a MgO unit cell. For larger thickness the signal decays exponentially with a decay length of \SI{1.8}{\AA}. Next, these experiments are repeated for HM=Ta. The results in \cref{fig:2}(d) surprisingly show an increase of the signal of more than \SI{20}\% for small thicknesses of MgO (below \SI{2}{\AA}). For larger thicknesses, again an exponential decay with a similar characteristic length scale as for HM=Pt is observed. We note that the signal amplitude vs. MgO thickness exhibits identical behavior when we employ broadband free-space THz field detection (see Fig. S1).

{\bf Calculation of spin currents for FM/MgO/HM trilayers}
In order to reveal the mechanism responsible for the enhanced ISHE signal in samples with MgO interlayer  and HM =Ta  ab-initio transport calculations are combined with density functional theory (DFT).   \cref{fig:2}(c) and (d) show the calculated MgO thickness dependence of the SC for a small bias voltage  of  50~mV. Clearly, for both Pt and Ta as HM the calculated SC  well reproduces the corresponding experimental ISHE data. For the case of Pt, the spin current (SC) decreases monotonously with the MgO thickness after showing a plateau behavior up to \SI{2}{\AA}. By contrast when Ta is used as HM, one first finds an increase of the SC up to 1 ML of MgO followed by a rapid decay well matching  the experimental results.

To understand the origin of the overall rapid decay of the SC with MgO thickness, we consider the projected local density of states (LDOS) for both systems, where the thickness $t$ of MgO barrier varies from 1 ML to 4 MLs. As seen in the extended data \cref{fig:S2}(b) and \cref{fig:S3} the MgO layer is actually metallic for a thickness of up to 2 MLs and only starts opening a band gap for thicknesses larger than 3 ML.  With increasing MgO barrier thickness the transport mechanism changes gradually from a metallic transport to a tunneling behavior and thus, as expected the SC drops exponentially with a decay length of 1.8 {\AA} as shown in \cref{fig:2}(c) and (d). 

\cref{fig:S2}(c) shows the spin- and energy-dependent transmision spectrum for the case of Fe/Ta and Fe/MgO(1ML)/Ta. The corresponding $\mathbf{k}$-dependent transmissions are presented in \cref{fig:S4} and \cref{fig:S5}. One finds that around the Fermi level the transmission is reduced by more than 50\% for both spin channels when 1~ML of MgO is inserted at the Fe/Ta interface. However, this reduction is more pronounced for the spin-down channel (minority-spin), which leads to an enhancement of the overall SC, i.e., the absolute value of the  $T^{\uparrow}-T^{\downarrow}$ at the Fermi level for Fe/MgO(1ML)/Ta is about 30\% larger than the corresponding value in Fe/Ta case [see \cref{fig:S2}(c)]. This accounts for the enhancement of the SC for HM=Ta case in \cref{fig:2}(d). This enhancement of the SC is linked to the recovery the magnetic moment of Fe(Co) and hence the spin polarization at the Fe/Ta (CoFeB/Ta) interface.

It is known that when elementary 3d ferromagnets are interfaced with transition metals with less than half filled 5d shell such as Hf, Ta, or W the magnetic moment at the interface is reduced \cite{Miura_aip_2013}. In some cases (e.g. Ni$_{81}$Fe$_{19}$) this effect can even give rise to formation of a magnetically dead layer \cite{NiFe_dead_layer}. This suppression of the magnetic moment can be qualitatively explained on the basis of the Stoner model by considering the density of states at the Fermi level $N(E_F)$  (see \cref{fig:S6} and \cref{fig:S8}(a)) and the Stoner parameter. Due to extended 5\textit{d}  orbitals of the early transition metals the strong Fe(3\textit{d})-Ta(5\textit{d}) hybridization at the interface transfers the Fe-3\textit{d} weight around the Fermi level to lower energies and thus $N(E_F)$ is substantially reduced. Therefore, the Stoner criterion for the Fe atoms at the interface is hardly satisfied. Note, that for simplicity only Fe atoms are considered in the calculations. However, we want to point out that for Co atoms at the interface the reduction of $N(E_F)$ is even larger than for the Fe case as illustrated in extended data figures \cref{fig:S7} and \cref{fig:S8}(b)). Thus the Co magnetic moments at the interface with Ta are also strongly suppressed. As one moves from the left to the right within the row of Elements in the Periodic Table, the nuclear charge of the HM increases causing the \textit{d}-wave functions to contract. This reduces the hybridization between Fe(3\textit{d})-Pt(5\textit{d}) orbitals and causes an enhancement of the Fe magnetic moments at the Fe/Pt interface.
\begin{figure*}[!]
\includegraphics[width=1\textwidth]{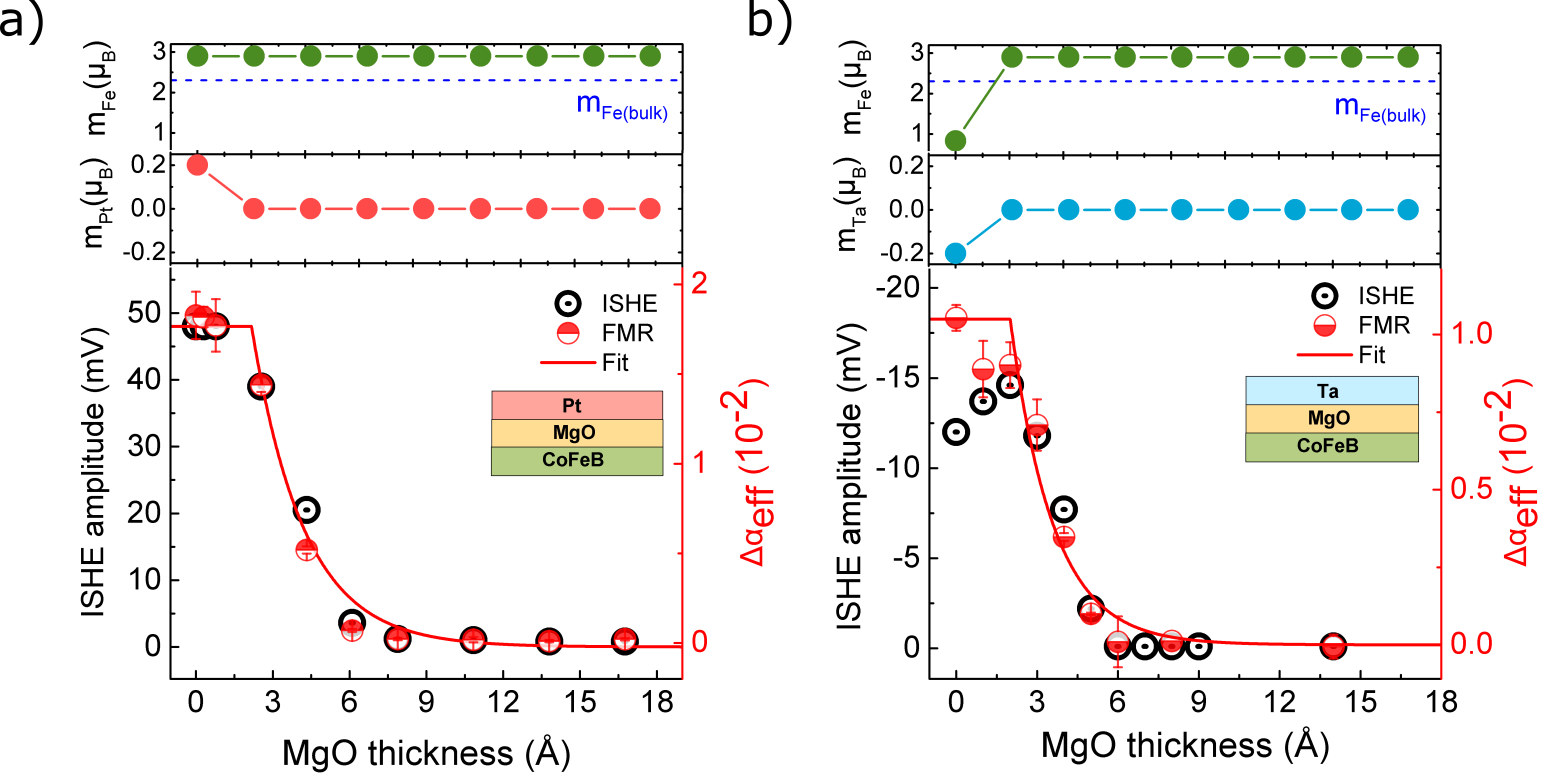}
\caption{\label{fig:3} {\bf ISHE signals, enhanced damping and interface magnetic moments}. MgO thickness dependence of the ISHE signal amplitude and the enhanced damping due to spin pumping for HM=Pt (a) and HM=Ta (b) as a heavy metal. The top panels of (a) and (b) show the calculated magnetic moment per atom at the interface for Fe and Pt (a) and Fe and Ta (b).
}
\end{figure*}
The calculated DFT values of interface magnetic moments are presented in \cref{table:1} and show that the spin magnetic moment of the Fe atoms at the interface layer are reduced to an average value of $\SI{0.8}{\mu_B}$ (see table S1), which only only corresponds to 30 percent of the Fe bulk magnetic moment. The reduced spin polarization at the interface layer also leads to a reduced spin current transmission. Interestingly, the insertion of 1 ML MgO at Fe/Ta junction causes the recovery of magnetic moment of Fe at the interface, resulting in an enhancement of the spin dependent transmission and SC in agreement with experimental data.    

{\bf Impact of the MgO interlayers on spin pumping}
To better understand the MgO interlayer dependence of the ISHE signal for HM=Ta layers we now consider the total spin current pumped out of  the FM layer. For this, we measure the additional damping caused by the MgO/HM capping layers due to spin pumping as shown in \cref{fig:S10}. This effect is sensitive to the total spin current emitted by the ferromagnet, while the electrically measured ISHE signal is only sensitive to the fraction of the spin current pumped into the HM layer. In \cref{fig:3}, we show the enhanced damping due to the proximity of the HM layer $\Delta\alpha_{\mathrm{FM/MgO/HM}}=\alpha_{\mathrm{FM/MgO/HM}}-\alpha_{\mathrm{FM/MgO}}$ as a function of the MgO interlayer thickness and compare it to the thickness dependence of the corresponding inverse SHE signal. For the case of HM=Pt, the ISHE signal and the additional damping the same MgO thickness dependence as shown in \cref{fig:3}(a). In contrast, for the HM=Ta case, the enhancement of the SC transmission for  thin MgO interlayers is not observed for the additional damping as shown in \cref{fig:3}(b). Instead the damping due to spin pumping reduces in a similar fashion as observed for HM=Pt. This implies, that part of the spin current pumped in the HM=Ta case is not delivered to the HM layer and lost at the interface. Such an effect is known as spin memory loss (SML) and can quantified by $\delta$ such as $J^{HM}_s=(1-\delta)J_s$ where $J^{HM}_s$ is the spin current on the HM side and $J_s$ the total spin current pumped from the FM side. In the case of HM=Pt, there is very close correspondence between the ISHE voltage and the enhanced damping due to spin pumping corresponding to $\delta_{\mathrm{Pt}} \approx 0$. For HM=Ta these results show that the magnetic moment recovery at the interface enhances the spin current transmission by about \SI{20}{\percent} and simultaneously suppress the SML effect. In fact, the magnitude of SML for Ta can be estimated to be $\delta_{\mathrm{Ta}} \approx 0.3$ without MgO interlayer and reduces to zero already with the insertion of 2 ML MgO (see \cref{fig:3}(b). We speculate that the SML for the FM/Ta interface may be caused by the antiparallel alignment of the induced Ta interface moment.

{\bf Discussion}
Inserting MgO interlayers between FM and HM layers suppresses the spin current transmission exponentially with a decay length of $\approx 1.8${\AA}. This effect can be well explained by the  calculated spin current transmission. By combing ISHE and spin pumping measurments as a function of MgO thickness we demonstrated a connection between reduced magnetic moments at the HM/FM interface and spin memory loss for HM=Ta.  Based on calculations of the electronic structure we conjecture that this effect occurs for all 5d heavy metals with less than half filled d-shells \cite{Miura_aip_2013}when interfaced with 3d ferromagnets. The insertion of an atomically thin MgO interlayer is sufficient to recover the interface of the ferromagnet. This is further supported by the calculated local magnetic anisotropic energy (MAE) favouring in-plane anisotropy at the interface of Fe/Ta where the magnetic moment is heavily reduced as shown in extended data Figure \cref{fig:S9}. 

In summary, we demonstrate that the orbital hybridization between FM and HM layers at the interface can lead to two effects which need to be avoided for efficient spin injection: 1. SML and 2. reduced spin polarization of the FM at the interface. As we show an MgO interlayer with a thickness of 2{\AA} leads to optimum results for the spin injection at Ta/CoFeB interfaces. We  believe that chemical control of the interface hybridization at the atomic scale (e.g. by ultrathin oxide layers) as demonstrated here is a promising approach to tune and enhance the interface spin transmission and thereby improve the efficiency of many spintronic devices.


\section {Methods}

{\bf Samples.} For the experiments, two sets of FM/HM bilayers with ultrathin MgO interlayers are perepared. The  layer stacks  have the following structure TaN(1.5)/CoFeB(2)/ MgO($t$)/Pt(4)/TaN(1.5) and TaN(1.5)/CoFeB(2)/MgO($t$)/Ta(3)/TaN(1.5), where the thicknesses of the individual layers are in nanometers.  The MgO thickness $t$ has been varied between 0 and 1.7 nm. All layers have been grown by DC Magnetron sputtering at pressure of 3 mTorr execpt for MgO grown by RF-sputtering using an off-axis gun tilted at a right angle of 90\degree from the substrate plane. The MgO layer grows crystalline on amorphous CoFeB as shown in Fig.S2 using transmission electron microscopy (TEM) micrograph confirming the (001) orientation of MgO. Atomic force microscopy has been performed on all samples and rms roughness is below 2{\AA} for all of them. We would like to point out that the insertion of the MgO interlayer has almost no measurable impact on the conductivity of the layer stack (see Fig.~S3). In particalyar this implies that the Ta layer remains in the highly resistive $\beta$-phase for all MgO interlayers.

{\bf Ultrafast inverse SHE.} Our samples are excited with an amplified Yb:KGW femtosecond laser system operating at \SI{1030}{\nano \meter}   wavelength with a \SI{300}{\femto \second} pulse width. On the samples the laser fluence is set to \SI{3}{mJ/cm^2} unless otherwise indicated. For signal detection, an rf-probe tip connected to a sampling oscilloscope (synchronized with the laser) is used to measure the ISHE voltage signals. The  bandwidth of the probe tip and the oscilloscope is limited to \SI{50}{\giga \hertz}. Therefore the measured signal is a convolution of the ultrafast ISHE signal with the response function of the rf circuit \cite{Hoppe_ACS_anm_2021}. An external magnetic field is set parallel to the sample plane and perpendicular to the ISHE voltage. The signal of interest is taken as the difference between the two voltages from the two opposite field orientations to eliminate other possible signal sources. 

{\bf Ferromagnetic resonance (Spin pumping)}
Samples, placed on top of a broad band coplanar waveguide and can be excited with rf field frequencies between 2 and 20 GHz. The field swept linewidth for the ferromagnetic resonance is extracted by fitting the experiemtal data to loretzian line shapes. The observed slope of the linewidth determines the Gilbert damping parameter. The enhancement of the Gilbert damping due to the presence of the HM layers is attributed to spin pumping and spin relaxation in the HM layer.

{\bf Transport calculations} To model our system,  we use interface builder in the QuantumATK package to construct a common unit cell for Fe/MgO/HM trilayers. The MgO matches well the bcc Fe in its (001) orientation where the oxygen atoms face directly the Fe atoms. We consider HM=Pt and HM=Ta. For the case of Pt, we consider it in its body centered tetragonal structure giving a mismatch of 3\% with bcc Fe.  As for the case of Ta, We build a $1 \times 3$ supercell of Ta (110) in its body centered cubic phase. The MgO has been varied until 6MLs and to account for effective thicknesses less than one ML, we consider intermixing between the first Fe and HM ML at the interface and MgO. A sketch of the device structure with 4 ML of MgO and Ta as a HM is shown in Fig.\,S3(a). Ground-state electronic structure calculations are carried out using DFT, implemented in the QuantumATK R-2020.09 package \cite{QuantumATK} with Perdew-Burke-Ernzerhof (PBE) parametrization of the generalized gradient approximation for the exchange-correlation (XC) functional \cite{perdew1996generalized}. We use PseudoDojo pseudopotentials \cite{QuantumATKPseudoDojo} and LCAO basis sets. A dense $20\times 20\times1$ ($20 \times 7 \times 1$)  $\mathbf{k}$-point grid for Pt case (Ta case) and a density mesh cutoff of 120 hartree are used. The total energy and forces converge to at least $\SI{1E-4}{\eV}$  and 0.01 eV/ $\si{\angstrom}$, respectively. The transport calculations are carried out using DFT combined with the nonequilibrium Green’s function  method (NEGF). We use a $20 \times20\times172$ ($20 \times7\times172$) $\mathbf{k}$-point grid for Pt and Ta case in self-consistent DFT-NEGF calculations. The current is calculated within a Landauer approach [57], where $ I(V) = \frac{2e}{h}\sum_{\sigma}\int \, T^{\sigma}(E,V)\left[f_{L}(E,V)-f_{R}(E,V)\right] \mathrm{d}E $. Here $V$ denotes the bias voltage, $T^\sigma (E,V)$ is the spin-dependent transmission coefficient for an electron with spin $\sigma$  and $f_L(E,V)$ and $f_R(E,V)$ are the Fermi-Dirac distributions for the left and right leads  which translates here to FM and HM, respectively. We  assume that the electronic system is thermalized and thus temperature effects on transport properties can be taken into account via the Fermi-Dirac distribution function. The transmission coefficient $T^\sigma (E,V)$ is calculated using a $100 \times 100$   ($100 \times 34$)   $\mathbf{k}$-point grid for the Pt and Ta cases. 
%

\renewcommand{\thepage}{S\arabic{page}}
\renewcommand{\thesection}{S\arabic{section}}
\renewcommand{\thetable}{S\arabic{table}}
\renewcommand{\thefigure}{S\arabic{figure}}


\onecolumngrid

\section {Supplementary Material}

\begin{figure*}[h]
\includegraphics[width=1\textwidth]{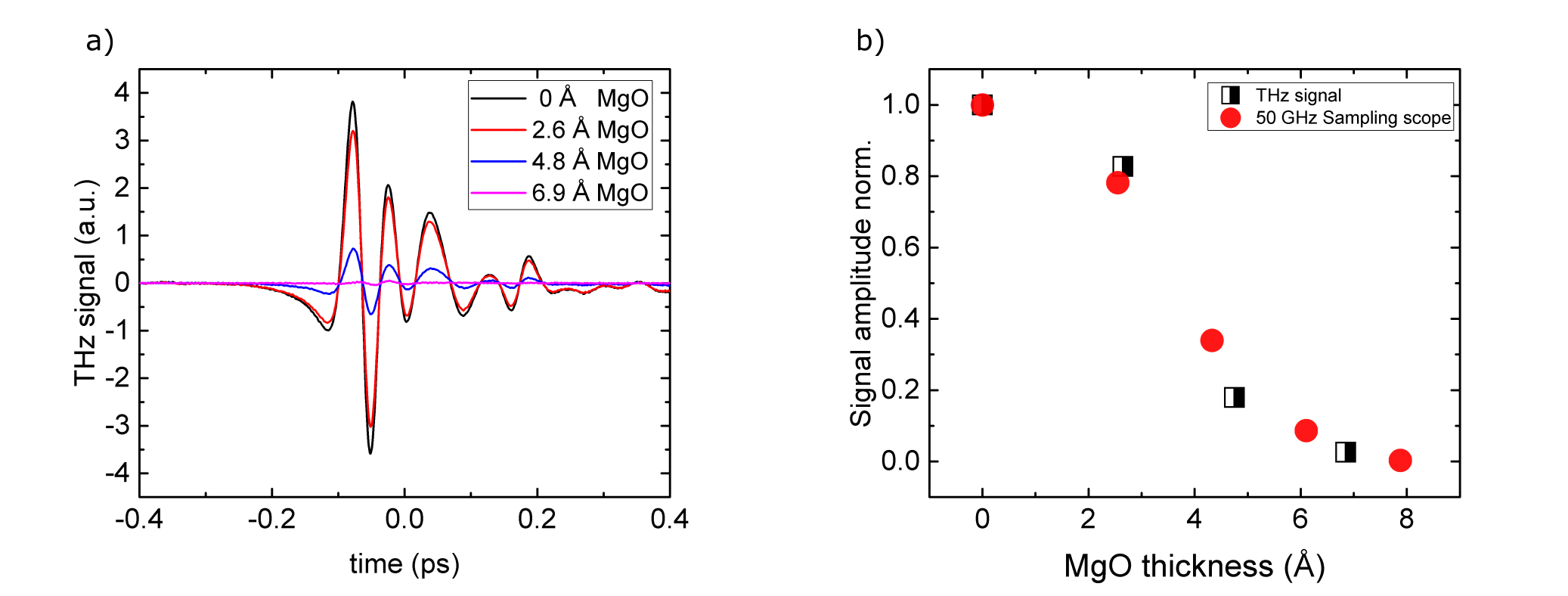}
\caption{\label{fig:S0} (a) Time resolved THz signals measured on CoFeB(2nm)/MgO/Pt(2nm) for different MgO thicknesses. (b) MgO thickness dependence of the rms value of the THz signals for CoFeB/MgO/Pt} 
\end{figure*}

\begin{figure*}[h]
\includegraphics[width=0.6\textwidth]{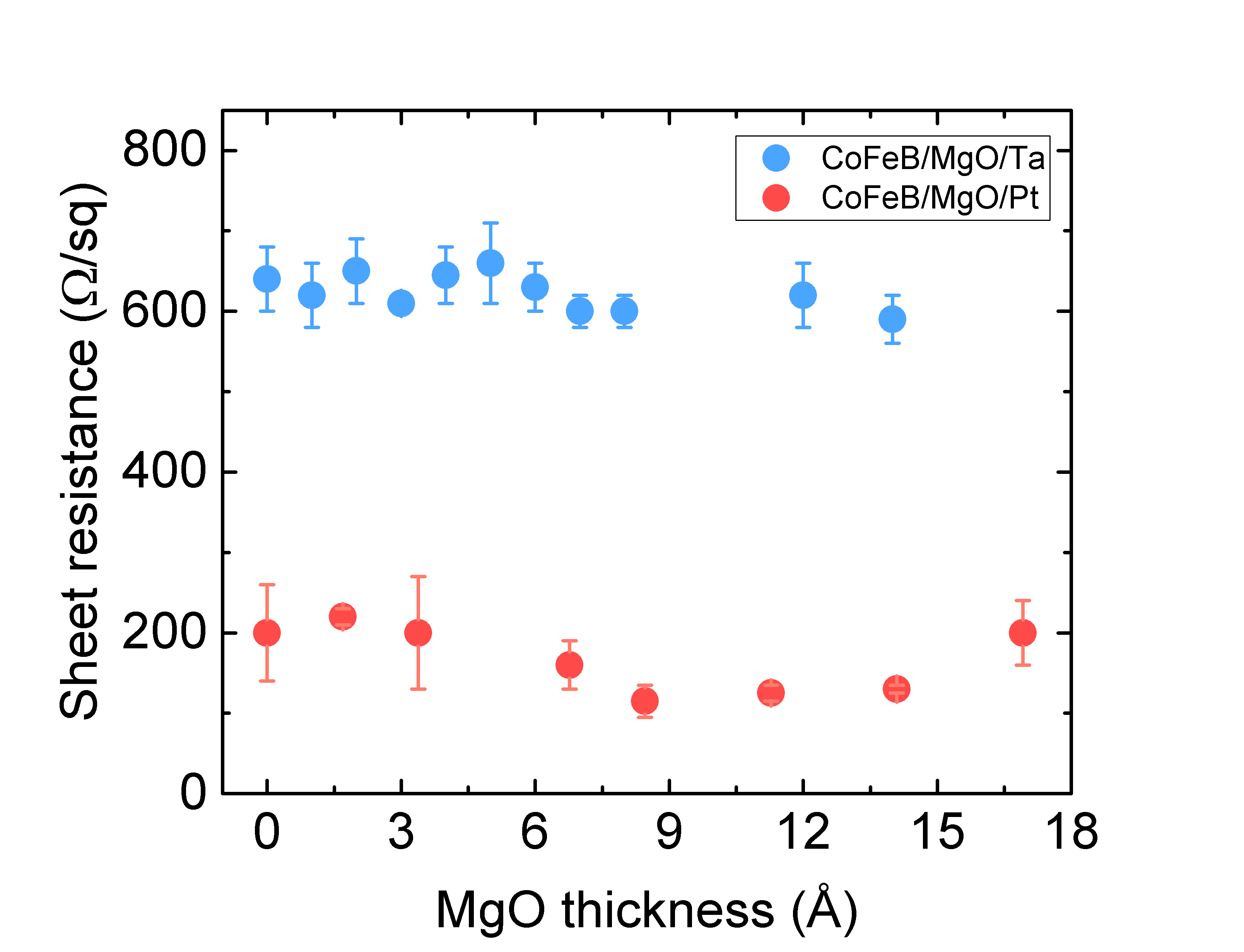}
\caption{\label{fig:S0.1} MgO thickness dependence of the sheet resistivity, measured with a 4 point probe method, for both cases where HM= Ta and HM=Pt. } 
\end{figure*}

\begin{figure*}[h]
\includegraphics[width=0.6\textwidth]{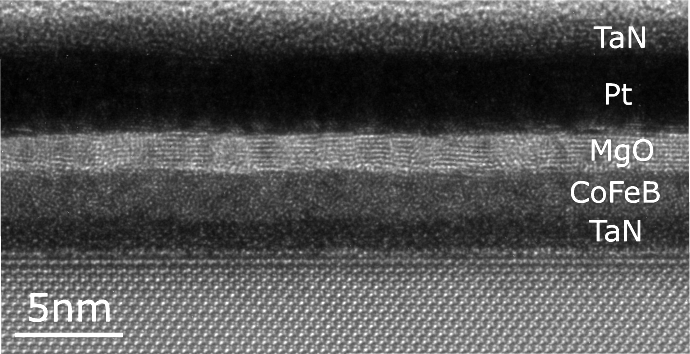}
\caption{\label{fig:S1} Transmission electron microscopy micrograph acquired at FEI TITAN 80-300 electron microscope showing the layered structure with Pt as a heavy metal and the MgO interlayer in its cubic phase with (001) orientation. From bottom to top: TaN(1.5), CoFeB(2), MgO(1.7), Pt(4), TaN(1.5). All thicknesses are in nm.} 
\end{figure*}

\begin{figure*}[h!]
\includegraphics[width=1\textwidth]{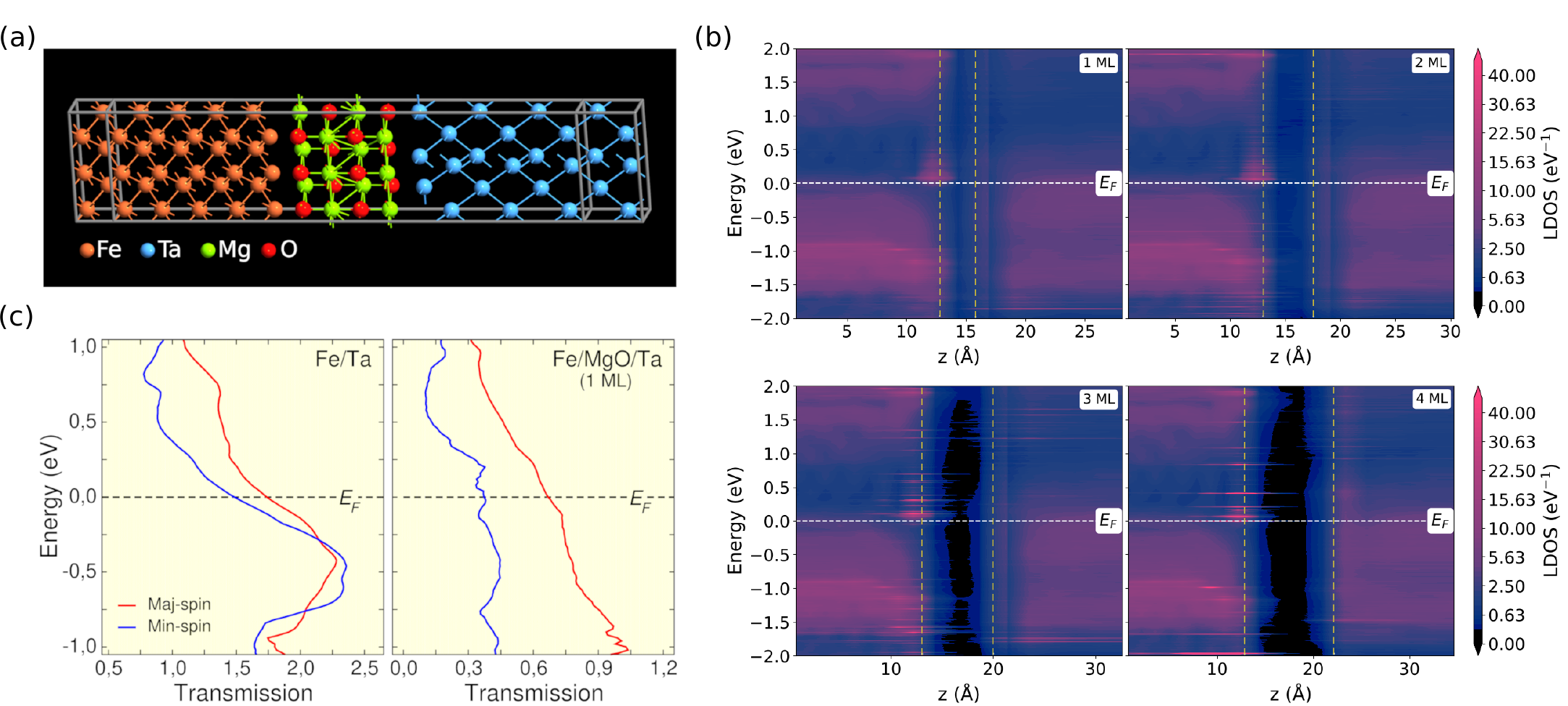}
\label{}
\caption{\label{fig:S2} (a) The atomic structure of the Fe/MgO/Ta junction. (b) The zero-bias  projected local density of states (LDOS) of the Fe/MgO/Ta junction  for different thickness of MgO barrier. The horizontal white dashed lines indicate the Fermi level. The vertical dashed lines denote the interface between Fe(Ta) and MgO. (c) Spin-resolved transmission spectra for Fe/Ta and  Fe/MgO/Ta junctions.} 
\end{figure*}

\begin{figure*}[h!]
\includegraphics[width=1\textwidth]{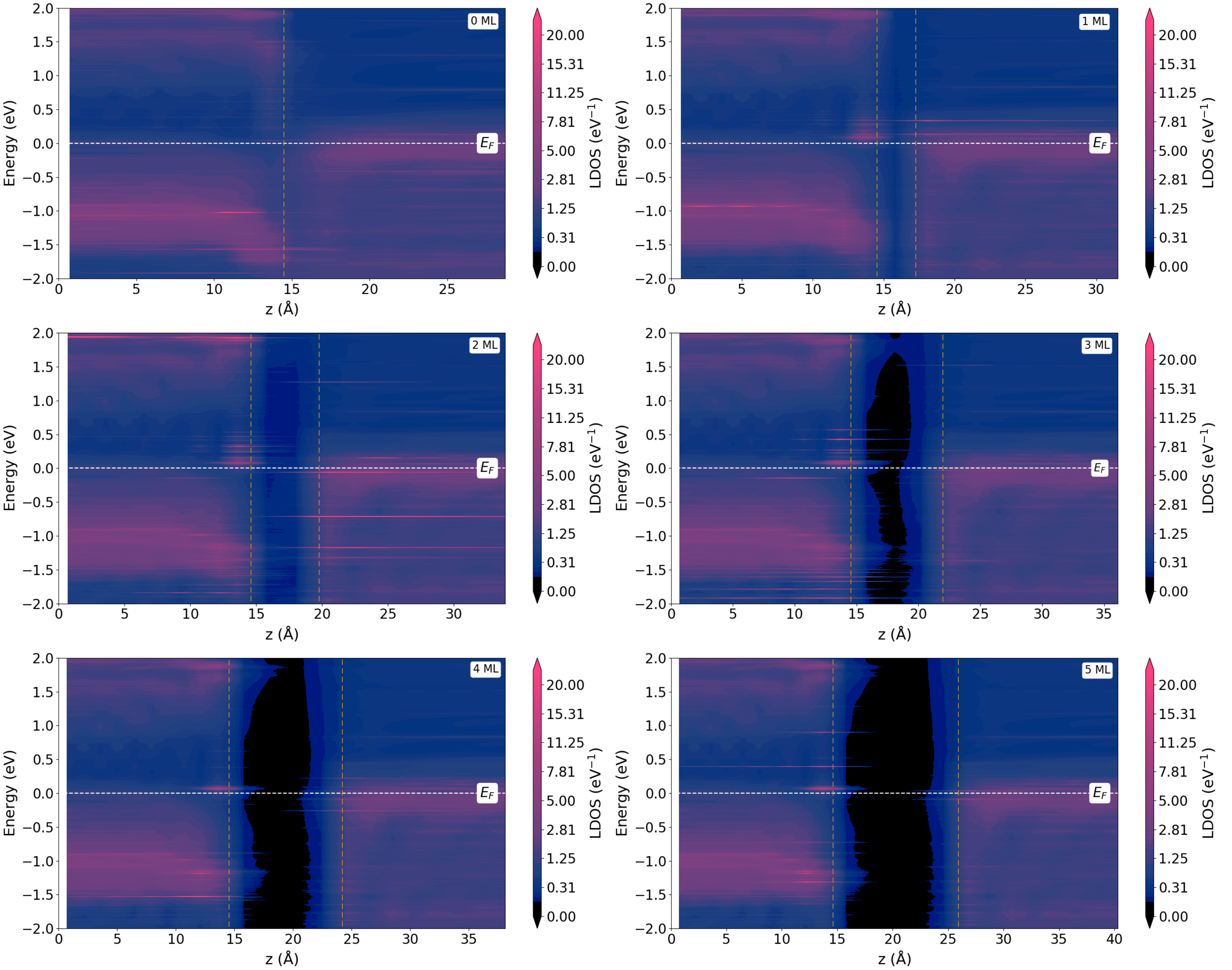}
\caption{\label{fig:S3} The zero-bias  projected local density of states (LDOS) of the Fe/MgO/Pt junction for different monolayers of MgO barrier. The horizontal white dashed lines indicate the Fermi level. The vertical dashed lines denote the interface between Fe(left), MgO(middle) and Pt(right).} 
\end{figure*}

\begin{figure*}[!ht]
\includegraphics[width=0.7\textwidth]{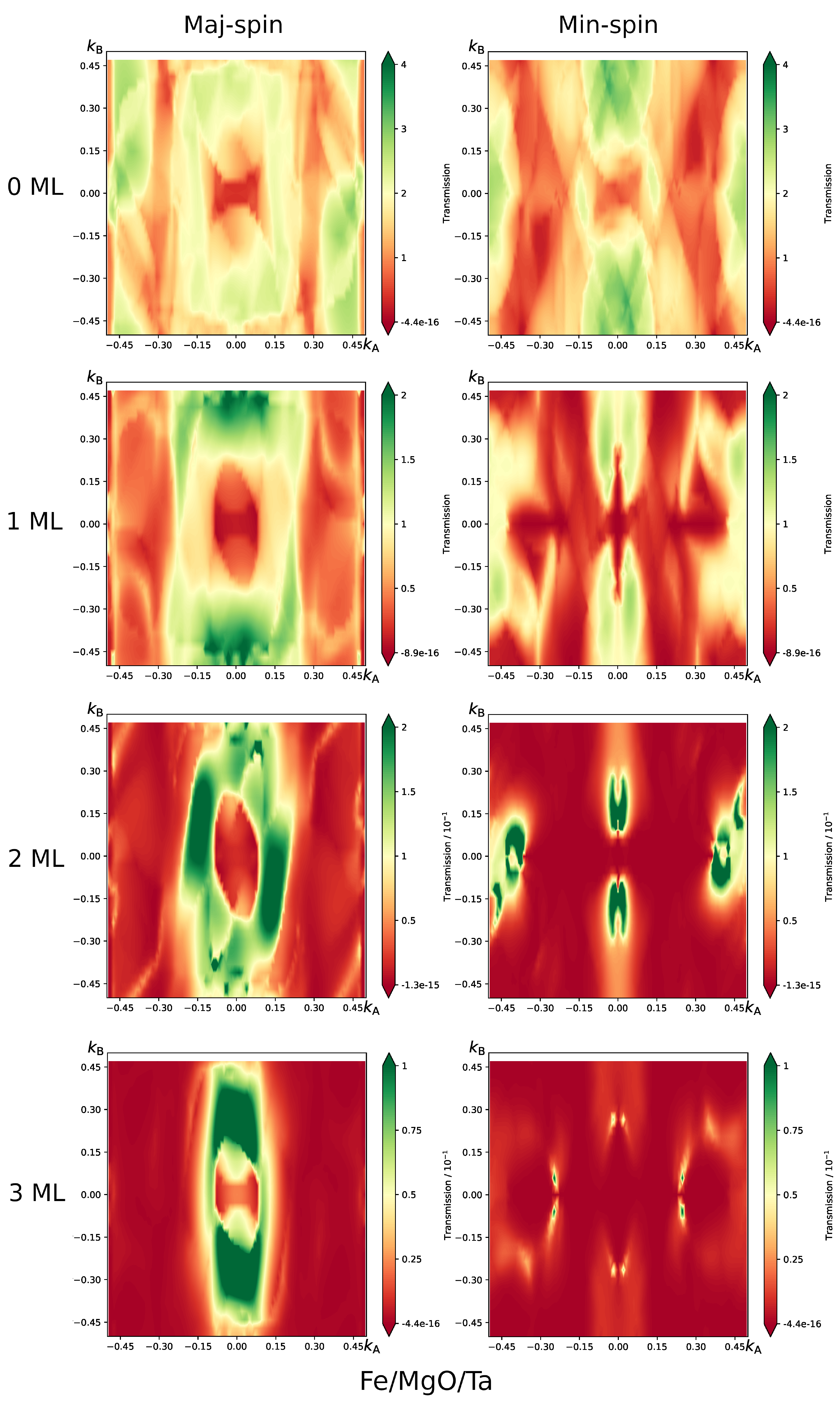}
\caption{\label{fig:S4} Spin resolved $\mathbf{k}$-dependent transmission amplitudes of the spin current through the entire Fe/Ta structure for different monolayers of MgO interlayer. } 
\end{figure*}

\begin{figure*}[h!t]
\includegraphics[width=0.7\textwidth]{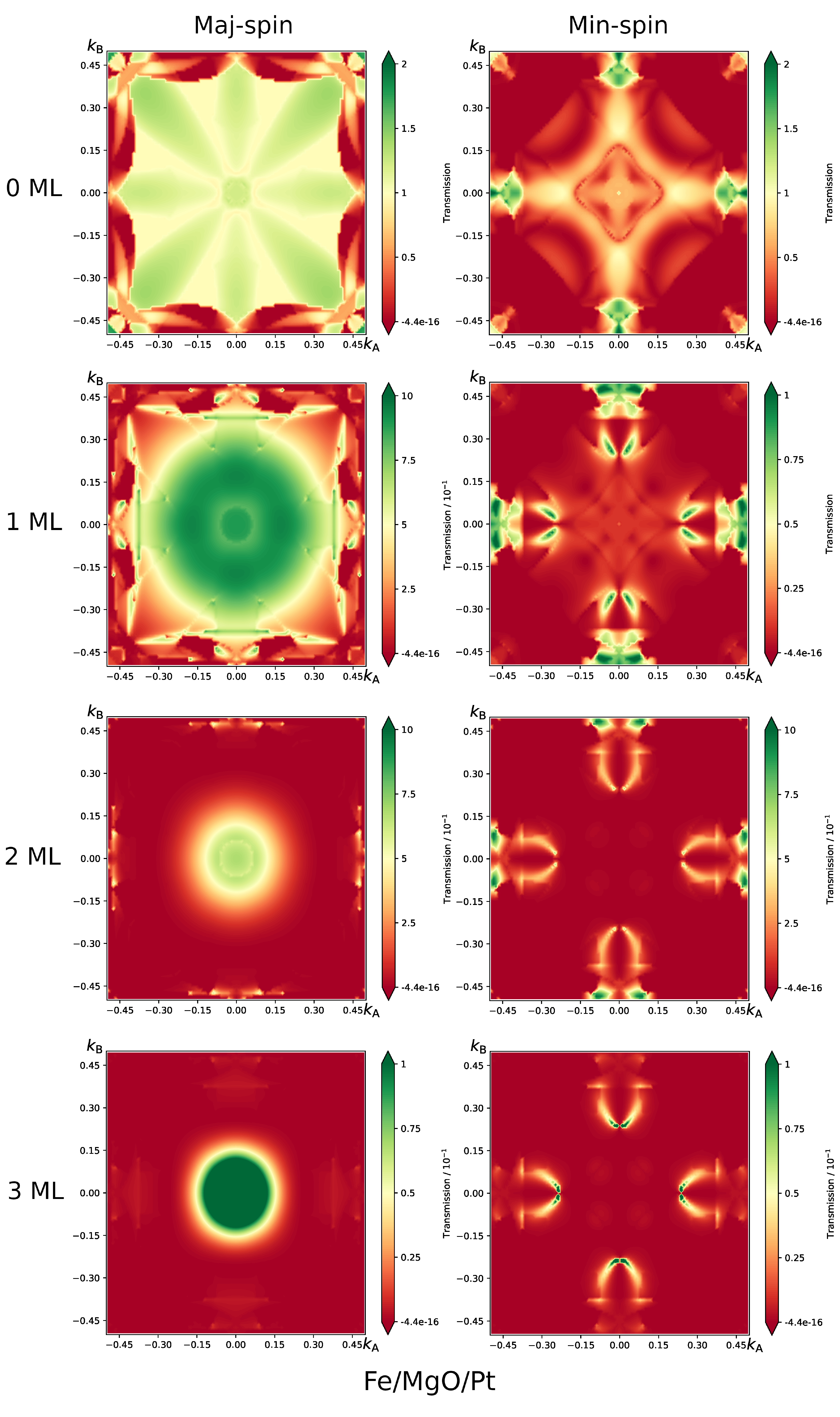}
\caption{\label{fig:S5} Spin resolved $\mathbf{k}$-dependent transmission amplitudes of the spin current through the entire Fe/Pt structure for different monolayers of MgO interlayer. } 
\end{figure*}

\subsection{Magnetic moment reduction}

\begin{figure*}[h!]
\includegraphics[width=1\textwidth]{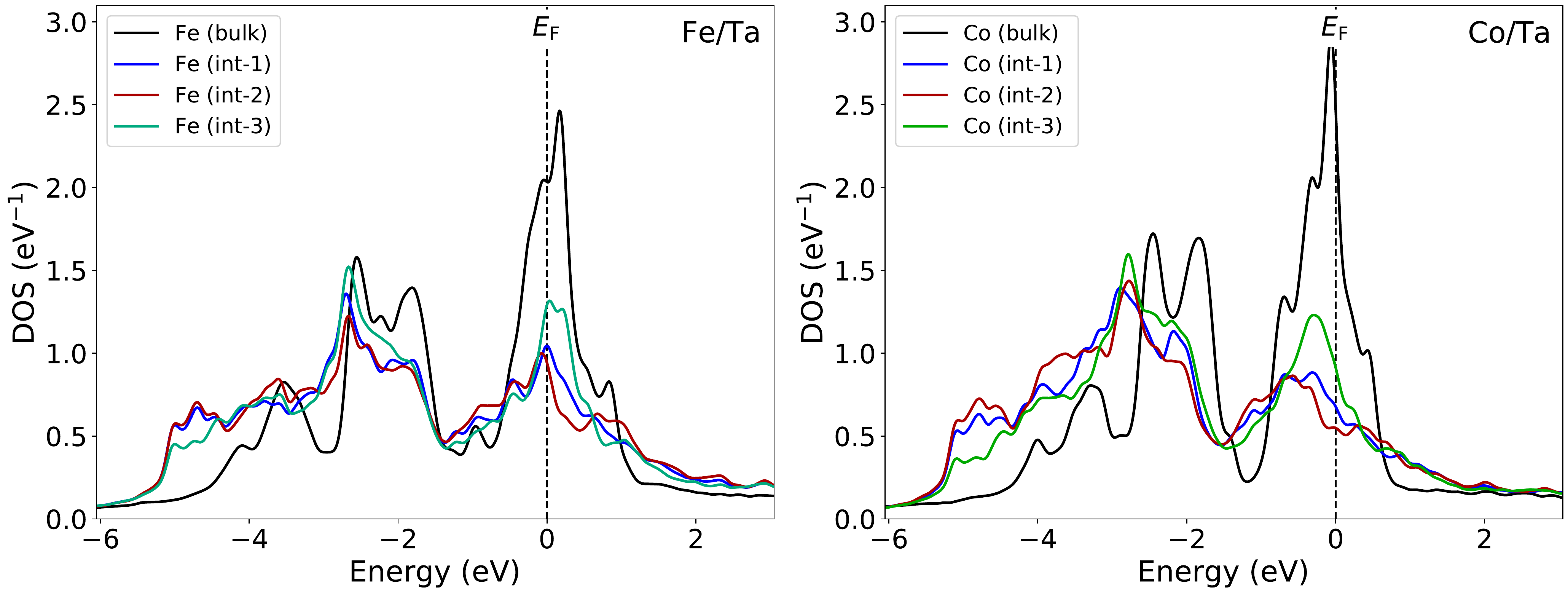}
\caption{\label{fig:S8} Non magnetic density of states for bulk Fe and the three interfacial Fe at Fe/Ta interface as well as for bulk Co and the three interfacial Co at Co/Ta interface.  } 
\end{figure*}

 In \cref{fig:S8} we present the non-magnetic DOS for the interface Fe (Co) atoms and compare with the corresponding bulk DOS. In Stoner model of ferromagnetism, the condition $I\cdot N(E_F)\geq 1$ should be satisfied, where $N(E_F)$ is the non-magnetic DOS at the Fermi level and the Stoner parameter $I$ can be related  to the Hubbard $U$ and Hund exchange $J$ parameter as it has been discussed in Ref\cite{heine_PRB_1990,sasioglu_PRB_2011}. We expect that the Stoner parameter $I$ would not change substantially for the interface Fe (Co) atoms since the $U$ and $J$ parameters are less sensitive to the changes in local environment \cite{sasioglu_PRL_2012} and thus the behavior of the DOS at the Fermi level play an essential role in suppression of the interface magnetic moments.
 
 \begin{table*}[h!]
\label{T:equipos}
\begin{center}
\begin{tabular}{ c | c || c | c || c |  c || c | c || c | c| }
\cline{2-10}
&& \multicolumn{2}{ c ||}{\textbf{Fe/Ta}}& \multicolumn{2}{ c ||}{\textbf{Fe/MgO/Ta}} & \multicolumn{2}{ c ||}{\textbf{Fe/Pt}}& \multicolumn{2}{ c| }{\textbf{Fe/MgO/Pt}}    \\ 
\cline{3-10}
& \textbf{Fe(B)}& \textbf{Fe} & \textbf{Ta} & \textbf{Fe} & \textbf{Ta}&\textbf{Fe} & \textbf{Pt} & \textbf{Fe} & \textbf{Pt} \\
\hline

\multicolumn{1}{|c|}{}&  &0.80& -0.16&2.92&-0.002&&&&\\
\multicolumn{1}{|c|}{magnetic moment ($\mu_{B}$)} & 2.3 &0.57& -0.22 &2.92&-0.03&2.92&0.2&2.89&-0.06\\
\multicolumn{1}{|c|}{}& &1.13& -0.21 &2.91&-0.03&&&&\\
\hline
\end{tabular}

\caption{Calculated magnetic moments for interfacial and bulk(B) layers in Fe/HM with and without one ML of MgO. For the case of Ta as a heavy metal, three atoms per layer are considered.}
\label{table:1}
\end{center}
\end{table*}

\begin{figure*}[h!]
\includegraphics[width=1\textwidth]{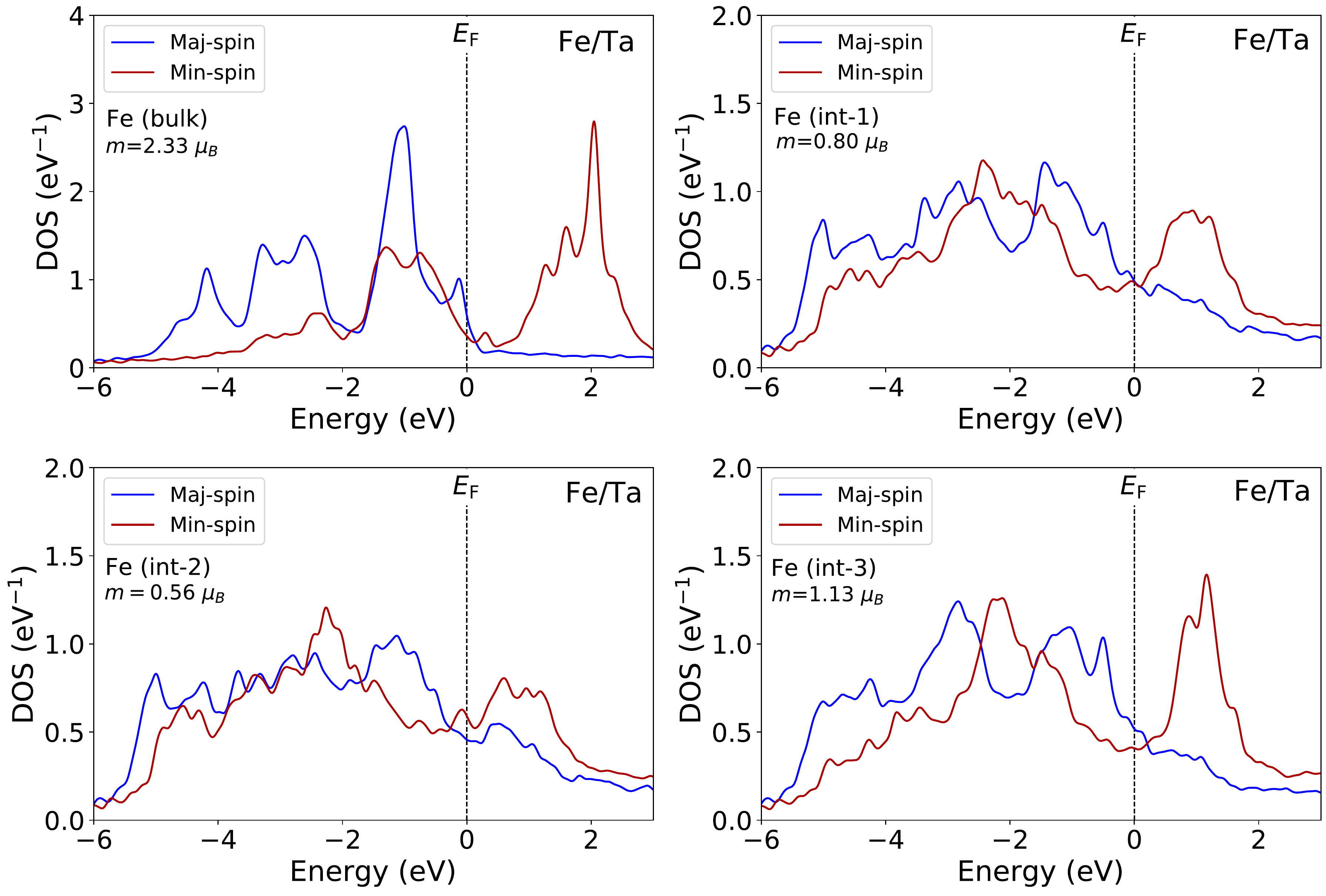}
\caption{\label{fig:S6} DOS for majority and minority spin for bulk Fe and the three different interfacial Fe at Fe/Ta interface. The magnetic moment is shown for all cases.} 
\end{figure*}

\begin{figure*}[h!]
\includegraphics[width=1\textwidth]{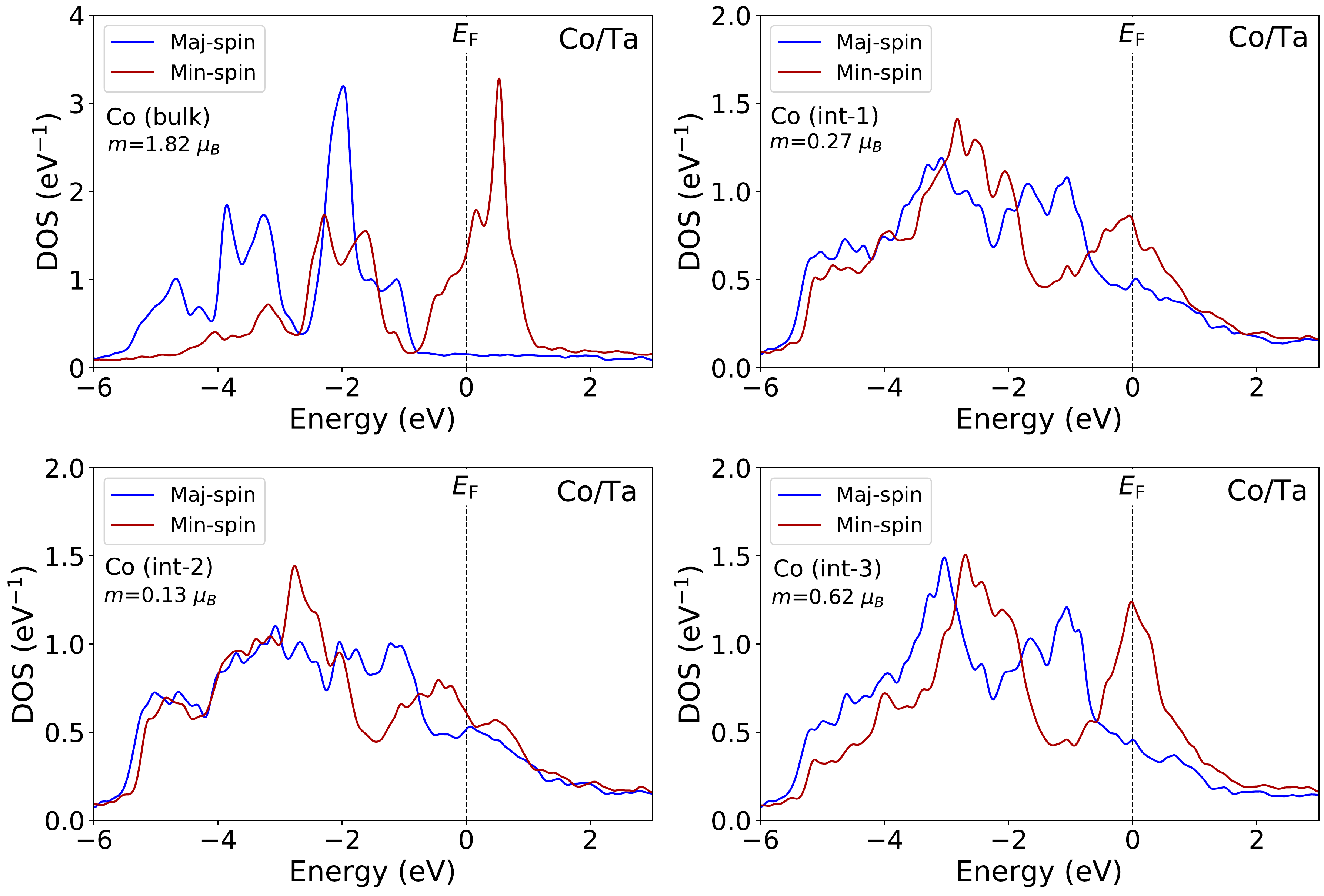}
\caption{\label{fig:S7} Density of states for majority and minority spin for bulk Co and the three different interfacial Co at Co/Ta interface. The magnetic moment is shown for all cases.  } 
\end{figure*}

\begin{figure*}[h]
\includegraphics[width=1\textwidth]{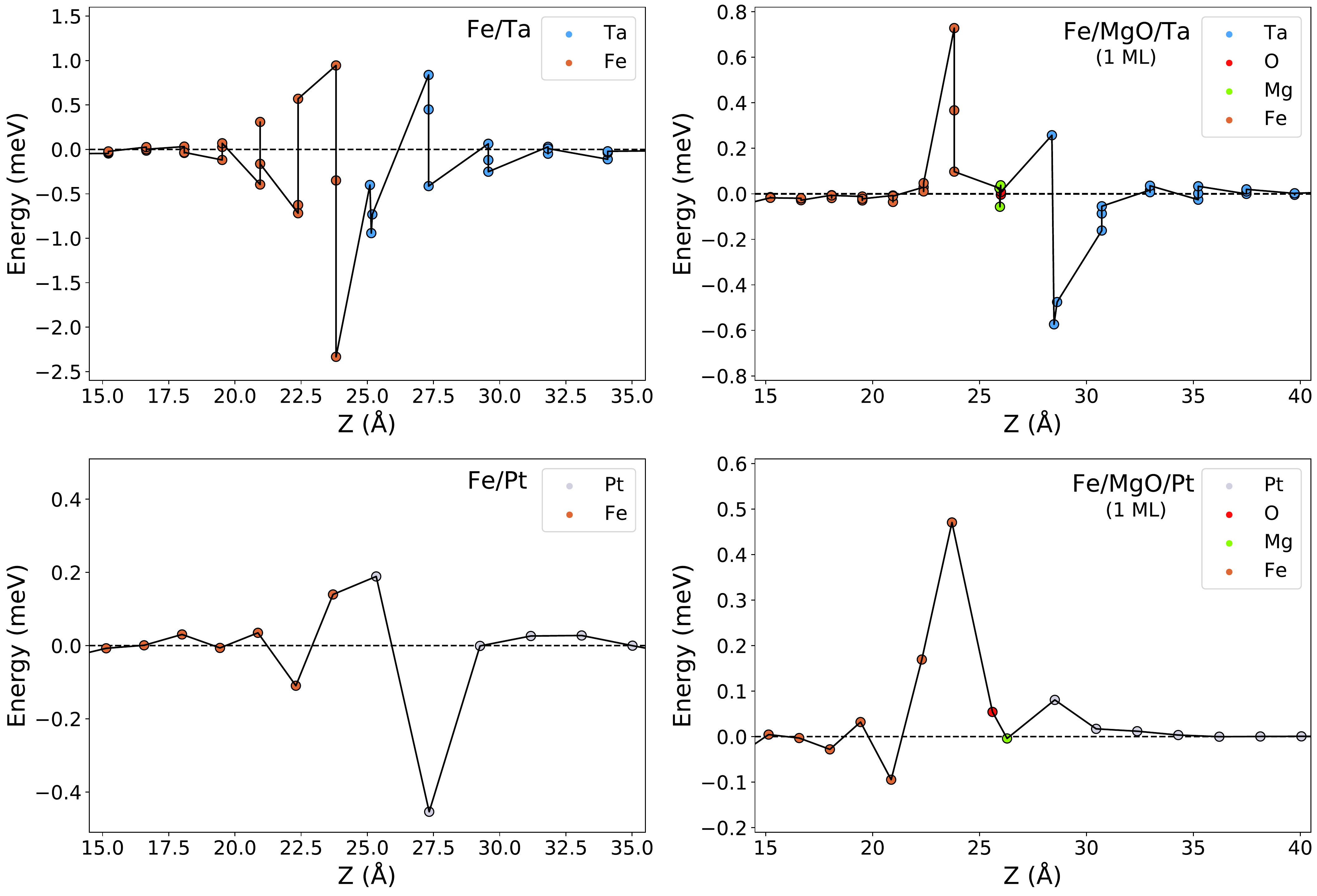}
\caption{\label{fig:S9} Local magnetic anisotropy energy for both heavy metals without MgO and with one ML of MgO.} 
\end{figure*}

\begin{figure*}[h]
\includegraphics[width=1\textwidth]{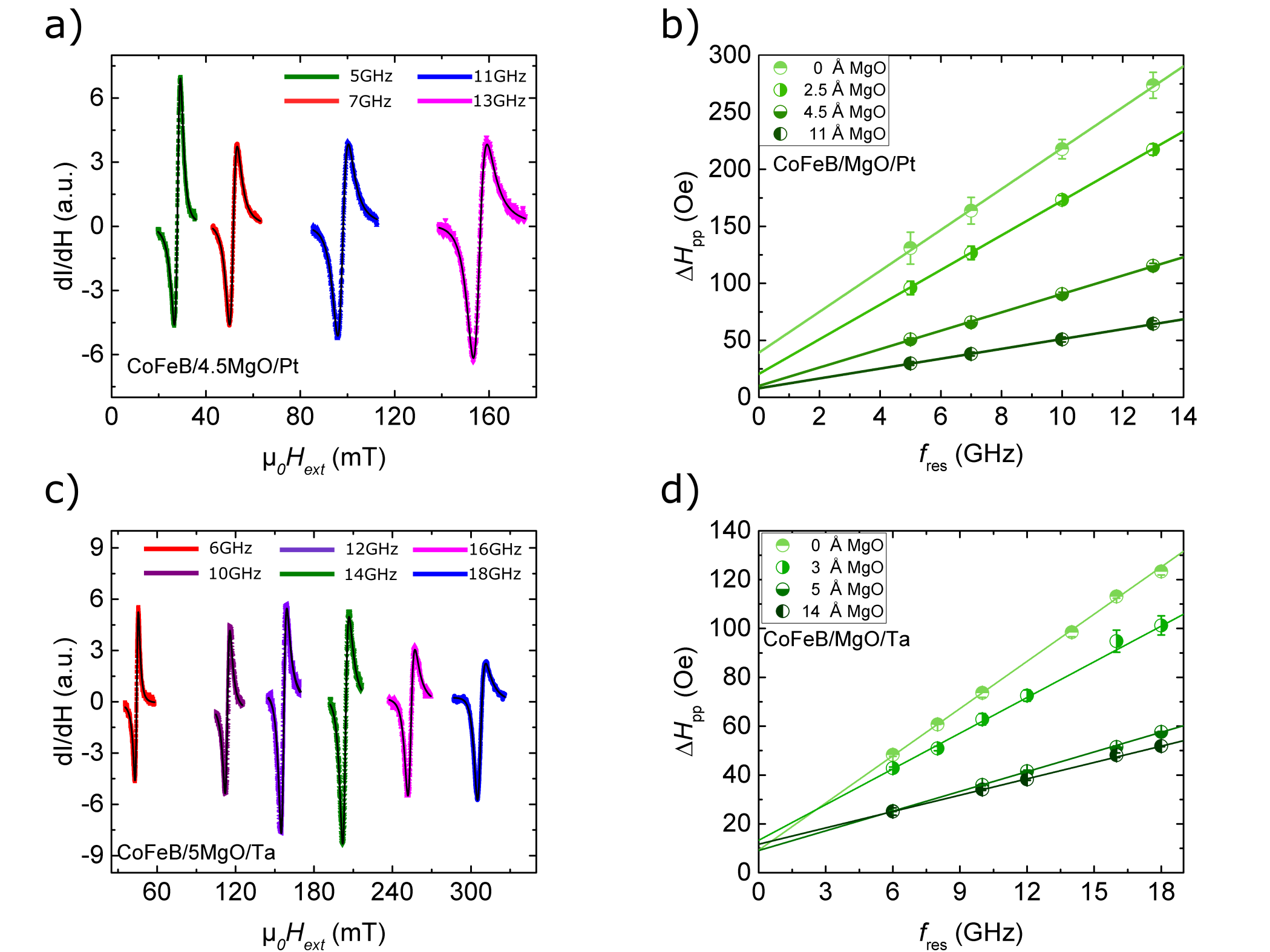}
\caption{\label{fig:S10} (a) and (c) Magnetic field dependence of the absorption spectra at the ferromagnetic resonance for various rf excitation frequencies for CoFeB/4.5MgO/Pt and CoFeB/5MgO/Ta respectively. (b) and (d) Frequency dependence of the ferromagnetic resonance linewidth for CoFeB/MgO/Pt and CoFeB/MgO/Ta respectively with various MgO thicknesses. } 
\end{figure*}



\end{document}